\documentclass[aps,prf,reprint,superscriptaddress,onecolumn]{revtex4-2}

\usepackage{graphicx}%
\usepackage{multirow}%
\usepackage{amsmath,amssymb,amsfonts}%
\usepackage{amsthm}%
\usepackage{xcolor}%

\usepackage{hyperref}
\hypersetup{
colorlinks=true,
linkcolor=blue,
anchorcolor=blue,
citecolor=blue}
\newcommand{\mdf}[1]{\textcolor{black}{{}#1}}

\begin{document}


\title{Unifying shear thinning behaviors of meso-scaled particle suspensions}


\author{Yuan Lin}
\affiliation{Institute of Ocean Engineering and Technology, Ocean College, Zhejiang University, Zhoushan 316021, China}

\author{Peiwen Lin}
\affiliation{Institute of Ocean Engineering and Technology, Ocean College, Zhejiang University, Zhoushan 316021, China}

\author{Yixuan Liang}
\affiliation{State Key Laboratory of Fluid Power and Mechatronic Systems and Department of Engineering Mechanics, Zhejiang University, Hangzhou 310027, China}

\author{Dingyi Pan}
\email[]{dpan@zju.edu.cn}
\affiliation{State Key Laboratory of Fluid Power and Mechatronic Systems and Department of Engineering Mechanics, Zhejiang University, Hangzhou 310027, China}

\date{\today}

\begin{abstract}
The rheology of suspensions with meso-scaled particles [with size of $O(10^2)\ \text{nm}$ to $O(10)\ \mu\text{m}$] is intriguing since significant non-Newtonian behaviors are widely observed although the thermal fluctuation (Brownain motion) of the meso-scaled particles is negligible. Here, we show that the linear constitutive relation for such systems fails due to a flow-induced particle aggregation, which originates from the inherent inter-particle interactions, e.g., the weakly adhesive van der Waals interaction. This accounts for the temporal evolution of the rheological property in both steady and oscillatory shear flows. A dimensionless number that measures the importance of the hydrodynamic interaction in shear flow with respect to the inter-particle interaction, \mdf{is} proposed, through which the non-linear constitutive relation for suspensions with various particle sizes, particle concentrations, as well as flow conditions could be unified. This investigation bridge \mdf{the gap between micro- and macro-scaled suspension systems} and make the rheology of the meso-scaled suspensions predictable.
\end{abstract}


\maketitle

\section{Introduction}\label{sec:intro}
\begin{figure}[htb]
\centering
  \includegraphics[width=0.6\textwidth]{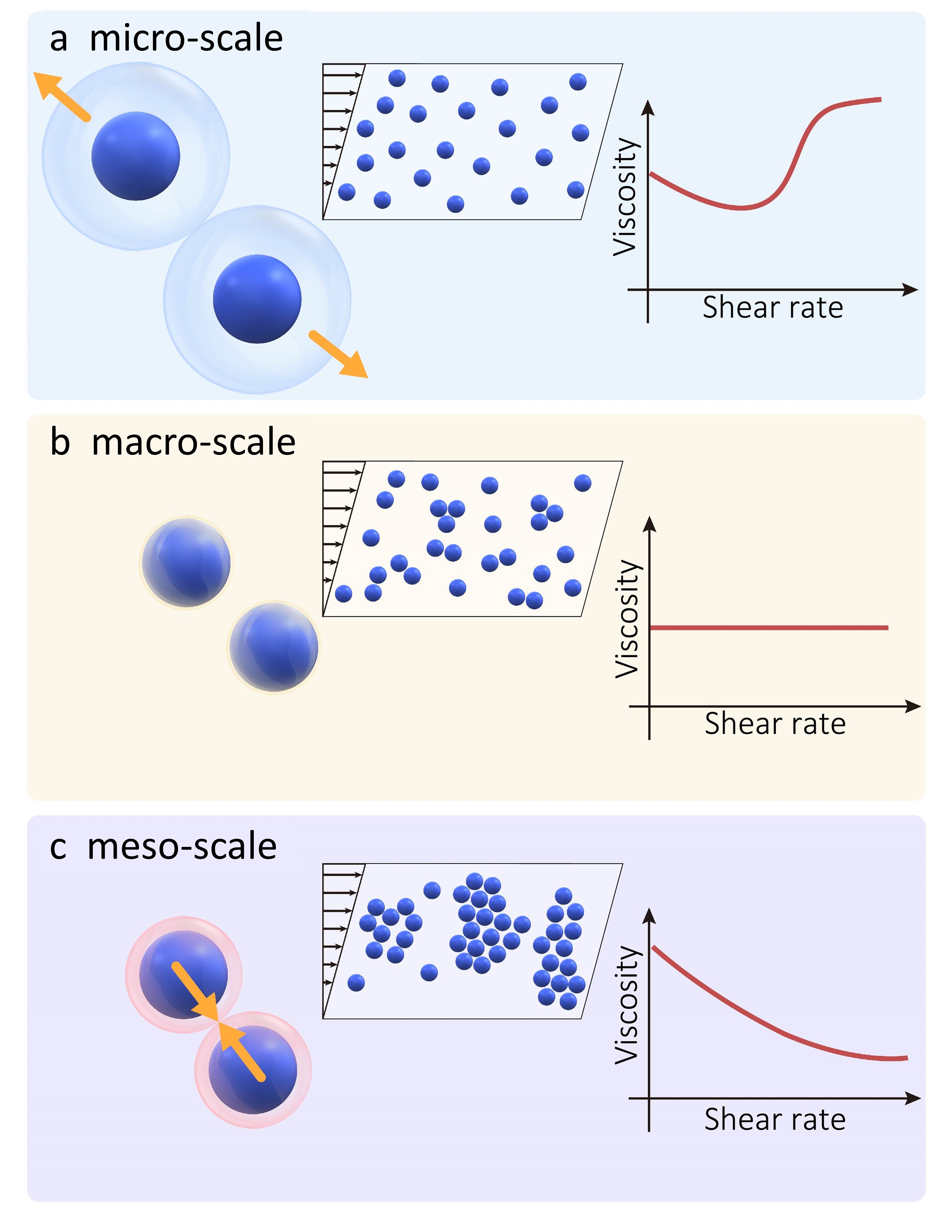}\\
  \caption{Diagram of the inter-particle interaction for neutral suspending particles with the size lies in different scales. (a) The interaction between {micro-scaled} particles is dominated by the Brownian motion, which acts like a strong repulsive inter-particle interaction that locks particles in a fixed position, forming cage-like structures~\cite{Foss2000}; (b) With increase of particle size to the {macro-scale}, all the surface-type particle interactions become negligible; (c) With the particle size lying in the {meso-scale}, \mdf{the near-field attractive interaction from the residual van der Waals interaction gives rise to shear-induced aggregation of particles}. The two insets in each figure show the resultant structure at low shear rates, as well as the corresponding rheological behavior, respectively.}\label{Fig0}
\end{figure}

For moderate and highly dense particle suspensions, inter-particle interactions are of great significance~\cite{Larson1999} -- they interplay with external hydrodynamic forces when the system deforms or flows, 
exhibiting intriguing bulk rheological behaviors and complicated local particle dynamics. 
As shown in Fig.~\ref{Fig0}(a), for suspension with micro-scaled particles, 
shear thinning due to 
particles' Brownian motion usually occurs 
and then followed by a shear thickening originated from particle aggregation based on a fore-aft asymmetric particle distribution
~\cite{Brady1997,Foss2000,Morris2009,Sierou2002,Guazzelli2018,Stickel2005}. 
Recently, it is found that increase of particle roughness promoted the shear thickening behavior, giving rise to a discontinuous shear thickening at a high particle loading~\cite{Blair_Ness_2022,Comtet2017,Safa2019}. 
For the simplest suspensions with {macroscopic} non-Brownian particles (particles too large to exhibit obvious thermal motion), the hydrodynamic interaction dominates, as shown in Fig.~\ref{Fig0}(b). Theoretically, the linearity of the Stokes equations suggests that the shear stress is linear with the shear rate~\cite{Guazzelli2018} and therefore it is Newtonian.

Nevertheless, the non-Newtonian behavior is prevailing in the experimental investigations on meso-scaled particle suspensions (particle size at the transition range, from {micro- to macro-scale, i.e., $ O(10^2) \;\text{nm}$ to $O(10) \;\mu\text{m}$}. 
For example, for suspensions with particle size of $O(10)\;\mu\text{m}$, notable shear thinning {is} observed~\cite{Singh2003,Dai2013,Lin2021a}; in the oscillatory shear with a small shear amplitude, a long-term irreversible evolution of the dynamic rheological properties {take} place although the shearing process is reversible~\cite{Bricker2006,Lin2013,Martone2021}. This unexpected shear thinning phenomenon {is} attributed to either shear-dependent deformation on the surface asperities~\cite{Chatte2018,Lobry2019,Arshad2021,Blanc2018} or particle aggregation~\cite{Papadopoulou2020,Gilbert2022,Richards2020}. The latter seems suitable to explain the transition of the viscosity and the consequent sharp shear thinning observed at a low particle volume fraction~\cite{Lin2021b}, and is also considered to be the origin for the long-term increase of the dynamic rheological behavior in the previously mentioned small amplitude oscillatory shear (SAOS)~\cite{Lin2015,Ge2021}. Upon a proper external impetus, e.g., a shear that cause particle displacement, rearrangement of particles could occur. However, considering the consensus that the hydrodynamic interaction dominates in non-Brownian particle suspensions, the key inter-particle interaction accounting for the aggregation process in such meso-scaled system remains largely to be understood.

In this work, we {unveil} the importance of surface-type inherent inter-particle forces, such as the weakly attractive van der Waals force, which {dominate} over the Brownian motion for meso-scaled particles, {on the particle organization and the consequent rheologcial behavior,} as shown in Fig.~\ref{Fig0}(c). This {promotes} particle aggregation in both steady and oscillatory shear flows. Our study explains the odd non-Newtonian behavior observed for the meso-scaled non-Brownian suspensions, through a dimensionless number ($\mathcal L$), that {measures} the importance of the hydrodynamic interaction with respect to the dominating inter-particle interaction. As a consequence, we {are} able to unify the rheological property (rate dependent shear viscosity) for the non-Brownian suspension system in both steady and oscillatory shear flows. 


\section{Results and discussion}\label{sec:results}
\subsection{Materials} 
For the meso-scaled particle suspension studied here, silicon oils (polydimethylsiloxane, Shin-Etsu Chemical Co., Ltd, Japan) with constant viscosity of $0.51 \; \rm Pa\cdot s$, $6\; \rm Pa\cdot s$, $10\; \rm Pa\cdot s$, and $30\; \rm Pa\cdot s$, respectively, are employed as the fluid medium. Polymethyl methacrylate (PMMA) particles with the density of $1.17\;\text{g}/\text{cm}^{3}$ are used as the suspending solid phase, while the silicone oil has the density of $\rho \approx 0.975\;\text{g}/\text{cm}^3$. Particles with {median diameters (D50) of $11.24\; \rm \mu m$, $69.11 \; \rm\mu m$ and $100.67 \; \rm\mu m$ 
respectively, are adopted to investigate the effects of particle size. In particular, for suspensions with particle diameter exceeding $50 \; \rm \mu m$, the silicon oils with the viscosity higher than $5 \;\rm Pa\cdot s$ are used to avoid particle settling at low shear rates due to density mismatch between the two phases. The dispersions are prepared by weighting a given amount of PMMA particles, and a given amount of silicone oil. Suspensions are stirred manually for $10 \;\rm mins$. All the samples are degassed in a vacuum oven for $24 \;\rm hrs$ to remove bubbles and trace-amount water that could influence the result~\cite{Lin2023}. This protocol produces reproducible samples which has been confirmed in the rheological experiment.

\subsection{Rheological Experiment}
A stress controlled DHR-1 rotational rheometer (TA Instruments) is used and the cone-plate geometry is adopted for small size particle ($10 \;\rm \mu m$) suspensions whereas the parallel-plate geometry is used for larger size particle suspensions to avoid unexpected particle--wall contact at the cone tip. 
A pre-shearing procedure with $\dot{\gamma} = 10\;\rm s^{-1}$ for $10 \;\rm s$ is applied to provide an identical initial state before the formal rheological test. The equilibrium viscosity curves 
are obtained through a logarithmic shear-rate sweep-down test and the shear rate range varies according to the tested system to avoid edge fracture at high shear rate. 
The shearing time in each step is set to $5 \;\rm s$ before the data is recorded. In the time sweep SAOS test, 
the strain controlled mode is adopted. The effect of the strain amplitude, $\gamma_{os}$, is investigated within the range from $0.1\%$ to $100\%$ and the oscillation frequency is set to $1 \;\rm Hz$. The effect of oscillation frequency is further investigated at $\gamma_{os} = 0.2\%$ with its magnitude changed from $0.5 \;\rm Hz$ to $5 \;\rm Hz$.

In our experiments of the meso-scaled suspension, notable temporal evolution of its rheological behavior is observed either in a steady or an oscillatory flow. As shown in Figs.~\ref{Fig1} (a) and (b), with a start-up steady shear (SS) and oscillatory shear (OS) applied to the suspension with {median particle diameter} of $11.24\;\mu\text{m}$, obvious rising could be observed for both the steady-shear and the complex viscosities, i.e., $\eta$ and $\eta^{\ast}$. Over time, the viscosity approaches to plateau values, $\eta_{\infty}$ and $\eta^{\ast}_{\infty}$, respectively. In SS, the transition process is controlled by the accumulated strain, $\gamma$, as shown in Fig.~\ref{Fig1}(a). $\eta_{\infty}$ is achieved when the applied $\gamma$ exceeds a critical strain, $\gamma_{c}$, which is independent to the shear rate. Shear thinning occurres as $\eta_{\infty}$ is shear-rate dependent. In the counterpart of OS, $\eta^{\ast}_{\infty}$ becames less obvious with increases of the strain amplitude, $\gamma_{os}$, and the oscillatory frequency, $\omega$ 
. Because of such unexpected long-term transition behavior in both SS and OS, it {is} commonly observed in the rheological test that the Cox--Merz rule~\cite{Cox1958} as $\eta(\dot{\gamma}) \approx \eta^{\ast}(\omega)$, with vanishing $\omega$ and $\dot{\gamma}$, which {is} successfully applied for polymeric systems, fails to hold for suspensions with meso-scaled particles~\cite{Lin2017}, since the equilibrium state is not reached in OS.

\begin{figure}[tb]
  \includegraphics[width=0.85\textwidth]{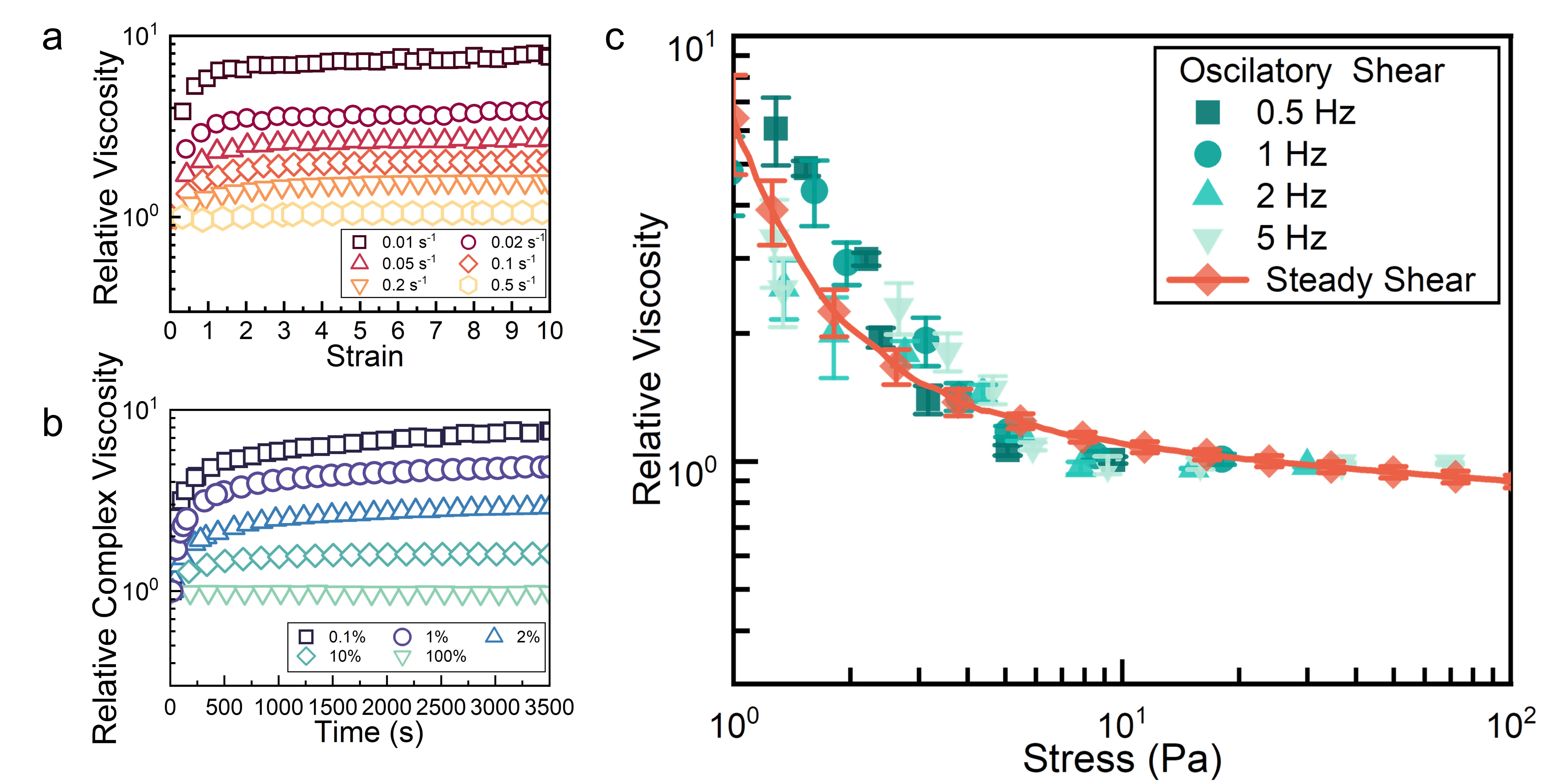}\\
  \caption{Temporal evolution of the (a) instant viscosity and (b) magnitude of the complex viscosity in steady and oscillatory shear, respectively, considering the variation of shear stress (or stress amplitude). (c) Change of the equilibrium viscosity with shear stress (or stress amplitude). The particle volume fraction of the suspension, $\phi = 40\%$.}\label{Fig1}
\end{figure}

Both the abnormal transition and the consequent shear thinning phenomenon {are} attributed to particle aggregation~\cite{Lin2013,Lin2021b,Bricker2006}. However, considering that the magnitude of current P\'eclet (${Pe}$) number is of $O(10^6)$, the leading interaction between the meso-scaled particles {is} the hydrodynamic response to the applied shear, by contrast, the particle Brownian motion that leads to the isotropic particle structure is negligible. Instead, an important adhesive-type inter-particle interaction which {is} comparable to the hydrodynamic interaction {is} needed to fully explain the aggregation behavior as conjectured in the experiment. On the other hand, as for {micro-scaled} suspension, in addition to the Brownian motion, the electrostatic and van der Waals inter-particle potential could be also significant~\cite{Larson1999,Wang2023}, which are
\begin{equation}
W_{e} = 2 \pi \varepsilon_0 \varepsilon a \psi_s^2 \text{ln}\left(\frac{1}{1-e^{-\kappa h_p}}\right),
\end{equation}
and
\begin{equation}
W_{vdw} = -A_H\frac{a}{12 h_p}\quad(h_p \ll a),
\end{equation}
respectively, where $h_p$ is the particle gap and $a$ is the particle radius. $\varepsilon_0$ and $\varepsilon$ are the dielectric permittivity of the free space and the dielectric constant of the medium, respectively. $\psi_s$ is the electrostatic potential at the particle surface. $\kappa^{-1}$ is the Debye length, and $A_H$ is the Hamaker constant. Obviously, all these inter-particle interactions decay with the increasing of particle size, and the range of interaction with respect to the particle radius, $h_{p{\infty}}/a$ (for electrostatic interaction, $1/ a \kappa$) also contracts. Furthermore, it is noted that the interaction from the Brownian motion ($\sim a^{-3}$) decays much faster than the van der Waals and electrostatic inter-particle interactions ($\sim a^{-1}$) as the particle size increases. Consequently, as the particle size reaches to meso-scale, we consider that the `weak' van der Waals interaction dominates for neutral suspending particles instead of the Brownian motion as the barrier that keeps particles away from each other, giving rise to a weak and near-field particle adhesion when particles are nearly in contact due to flow. This also explains the transition of the viscosity before equilibrium, as shown in Fig.~\ref{Fig1} -- instead of the immediate building up of the resistance in colloidal suspensions due to strong DLVO interaction or the Brownian motion, the meso-scaled particles needs to be first drived together in shear as a result of the acting of the weak particle adhesion. Furthermore, due to the absence of the thermal fluctuation at meso-scale, the flow-induced particle aggregation is considered to be stable when the flow is ceased. This is confirmed by previous studies -- instead of the above mentioned transition behavior, the stress built up immediately if the start-up shear is applied for a second time~\cite{Narumi2002}; in an amplitude sweep test following a start-up shear procedure with a low shear rate (at the shear thinning range), a considerably {higher} plateau modulus could be measured at $\gamma_{os} \leq 1$ before the structure was broken by the large-amplitude oscillation. 

\subsection{Numerical Simulations}
To validate the importance of the van der Waals interaction, a three-dimensional numerical simulation is performed with the Lubrication Dynamics--Discrete Element Method (LD-DEM). We assume the system to be simple by neglecting the Brownian motion of particles. Three typical interactions, namely, the hydrodynamic force, $F^H$, the contact force, $F^C$, and the van der Waals attractive force, $F^A$, are included in the meso-scaled system. The pairwise short-range hydrodynamic interaction includes Stokes' drag and lubrication forces, the details can be found in those published papers~\cite{cheal2018a,Wang2023}. The contact interaction is composed of normal contact force $F^C_n=k_n\delta$ and tangential friction force $F^C_t=k_t u$, in which, $\delta$ denoted the normal overlap between contact particle, $u$ denoted the tangential increment displacement. The Coulomb friction law $|k_tu|\le \mu k_n \delta$ is satisfied, where $\mu$ is the friction coefficient and it is set to 1, $k_n$ and $k_t$ are normal and tangential spring constants. The van der Waals attractive force {is} defined as $F^A=A_H^S a/12(h_p+\varepsilon')^2$, where $A_H^S$ {represents} the Hamaker constant in simulation system, $\varepsilon'$ {is} the particle roughness that determines the minimal particle separation. The simulation of this particle system is implemented with LAMMPS. We simulate $2000$ bidispersed particles with equal numbers of particles with radii $a$ and $1.4a$ and set the volume fraction to $40\%$, which is the same as in the experiments. In the initial state, the particles are randomly placed in the simulation box, with the rate-control shear progressed, the suspension flows as the three-dimension periodic box deforms, the shear stress and particle structure gradually reach to equilibrium. 

Typical simulation result {is} shown in Fig.~\ref{Fig2}, where the in-contact particle pair {is} indicated by a center-to-center thread. As shown in Fig.~\ref{Fig2} (a), in a start-up shear flow, fraction of in-contact particle pairs {increase} with the increasing of applied strain to the system, indicating the formation of clusters. At the equilibrium state, as shown in Fig.~\ref{Fig2}(b), the scale of particle clusters {decreases} as the shear rate increases. This {confirms} the notable {structure formation} at low shear rates for meso-scaled particle suspensions due to the presence of the adhesive interaction such as the weak van der Waals attractive interaction. Shear viscosity of the simulated suspension system is also calculated, as shown in Fig.~\ref{Fig3}(b), which presents a clear shear thinning behavior due to particle aggregation.

\begin{figure}[tb]
  \includegraphics[width=0.85\textwidth]{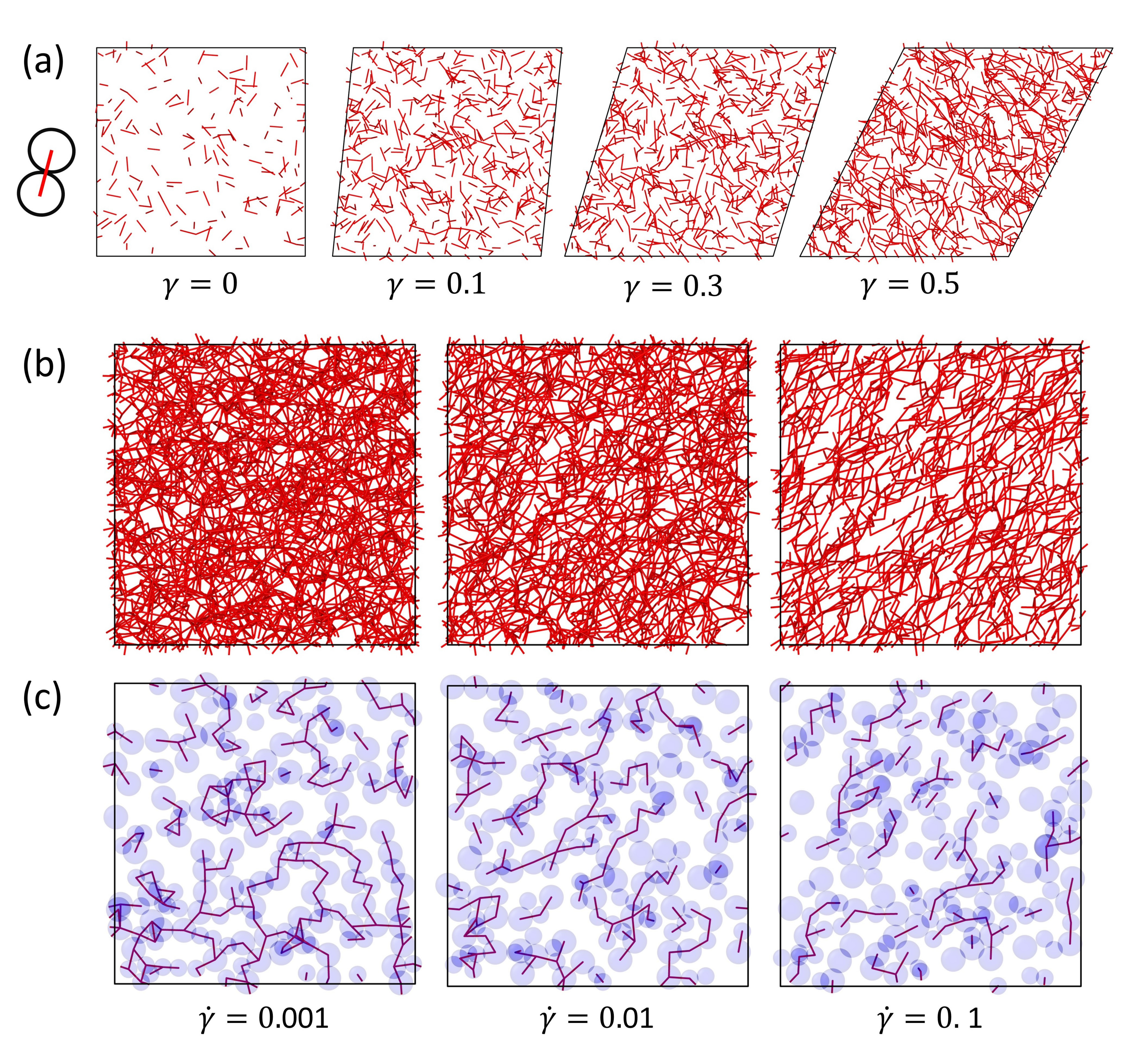}\\
  \caption{Particle structure investigated through numerical simulation in (a) a start-up shear flow at $\dot{\gamma} = 0.001 s^{-1}$; (b) at the equilibrium state in shear flow with various $\dot{\gamma}$. (c) structures in the cross section at the center of (b). Contacts between particles were represented by the center-to-center threads in the figure.}\label{Fig2}
\end{figure}


\subsection{Discussions}
This mechanism well explains the significant divergence observed on the shear thinning between different studies on this specific meso-scaled systems. The van der Waals interaction at the meso-scale is sensitive to the particle size and roughness. The adhesive interaction from the van der Waals interaction decays with increasing size of suspending particles. Therefore, as shown in Fig.~\ref{Fig3}(a), with further increment of the particle size, shear thinning {become} less obvious. It could be inferred that with particle size of $O(10^2)\;\mu\text{m}$, the system {become} nearly Newtonian. Also, shear thinning {is} found to be suppressed with increasing particle roughness~\cite{Lin2021b} -- the surface asperities {segregate} approaching particles, which {reduces} the adhesive interaction and consequently the aggregation behavior. Moreover, apart from the van der Waals force, interaction between suspending particles of meso-scale level could be easily altered by a small amount of third-phase droplets dispersed in the suspension~\cite{Lin2023,Koos2011,Koos2012}. The fine droplets {would induce} additional {near-field} interactions between the solid particles, enhancing {or suppressing} the particle aggregation and the consequent shear thinning behavior. This indicates that the weak near-field inter-particle interaction is sufficient to considerably alter the flow behavior of the meso-scaled suspension system.

Based on the above discussion, we {are} able to unify the ``shear thinning" observed both in OS and SS. As shown in Fig.~\ref{Fig1} (c), the variations of the normalized viscosities, $\eta^{\ast}_r=\eta^{\ast}/\eta^{\ast}_0$ and $\eta_r=\eta/\eta_0$, at the equilibrium state can be simply unified by scaling the data with the shear stress, $\tau_{ss}$, in SS or the shear stress amplitude, $\tau_{os}$, in OS, respectively, where $\eta_0$ and  $\eta^{\ast}_0$ {are} the steady-shear and complex viscosities at the initial state ($t = 0\;\text{s}$). This {gives} rise to $\eta_r(\tau_{ss}) \approx \eta^{\ast}_r(\tau_{os})$ at the equilibrium state, reproducing the form of Cox--Merz rule based on the stress instead of the rate for polymeric systems. The shear-induced aggregation structures {are} similar in both OS and SS, which {are} solely determined by the stress (amplitude) applied to the system. Given a proper driven to the system with a determined shear stress, meso-scaled particles reorganize to form a unique structure regardless the form of the shear. We {consider} that the applied shear flow (either OS or SS) {has} its effect on the suspension in two folds: First, it {promotes} particle aggregation, giving rise to the temporal elevation of the viscosity; Second, it {confines} the scale of the ordering structure at the equilibrium. As can be inferred in Fig.~\ref{Fig1} (c), a complete particle aggregation occurs only at a vanishing stress.


In order to unify the shear thinning behaviors of meso-scale suspensions, and meanwhile take account of the particle size, a dimensionless number to measure the importance of the inter-particle force with respect to the hydrodynamic force {is} proposed, 
\begin{equation}
\mathcal L(a) = \tau / \tau_p.
\end{equation}
Here, $\tau_p \approx C \left\langle F_p \right\rangle/\pi a^2$ {estimates} the ensemble stress from the inter-particle interaction, $F_p$. $\tau$ is the shear stress or stress amplitude, $\tau_{ss}$ or $\tau_{os}$, applied to the system and $C$ is a scaling parameter. With $\left\langle F_p\right\rangle = kT/6a$ that estimates the inter-particle interaction from the Brownian motion, the $\mathcal{L}(a)$ number {reproduces} the form of the P\'eclet number. Since the suspension system in this study {is} neutral, the $\mathcal{L}(a)$ number {is} estimated solely based on the van der Waals force, $F_{vdw}$. Consequently, 
\begin{equation}
\tau_p \approx C \frac{\left\langle F_{vdw}\right\rangle}{\pi a^2} =C A_{H}\frac{1}{12\pi a {h_p}^2}.
\end{equation}
It {is} proposed that {$h_p \approx \epsilon'$}, where $\epsilon'$ is the particle roughness considering that in contact, particles {are} separated by aspires on the particle surface. For the untreated and smooth particle, the roughness is {set to be linear with} the {radius} of the particle {(e.g., $\epsilon'= 10^{-3}a$)}. Consequently, if we take {$C = 1$}, then
\begin{equation}
{\tau_{p} \approx A_{H} \frac{1}{12 \pi a^3}\times 10^8 .}
\end{equation}

{For our suspension system, we calculate $A_H \approx 2.0449 \times 10^{-21}$ 
. As shown in Fig.~\ref{Fig3} (b), we {re-scale} the viscosity curves (normalized by the constant viscosity achieved at high shear rates, $\eta_{\infty}$) for suspensions with various particle sizes using $\mathcal{L}(a)$ number. It shows that viscosity curves could be reasonably unified, as shown in Fig.~\ref{Fig3} (b), although particles {are} not ideally mono-dispersed, and the particle roughness {is} roughly estimated. {Meanwhile, the simulation results in Fig.~\ref{Fig3} (b) show that, by using the dimensionless parameter $\mathcal{L}$, the viscosity curves with different attractive strengths merge into one main curve, indicating that van der Waals interaction dominates the shear thinning.} {The master shear thinning curve from the experiment approaches the one from simulation, if the particle gap to rescale the experimental data is set as $h_p = 10^{-4}a$ being lower than $h_p = \epsilon' = 10^{-3}a$ for common smooth particles. It is considered that, virtually, aspires on particles at such low-roughness level could not ideally separate particles as is proposed in the simulation, leading to a considerably higher $\tau_p$.} 


\begin{figure}[tb]
  \includegraphics[width=0.8\textwidth]{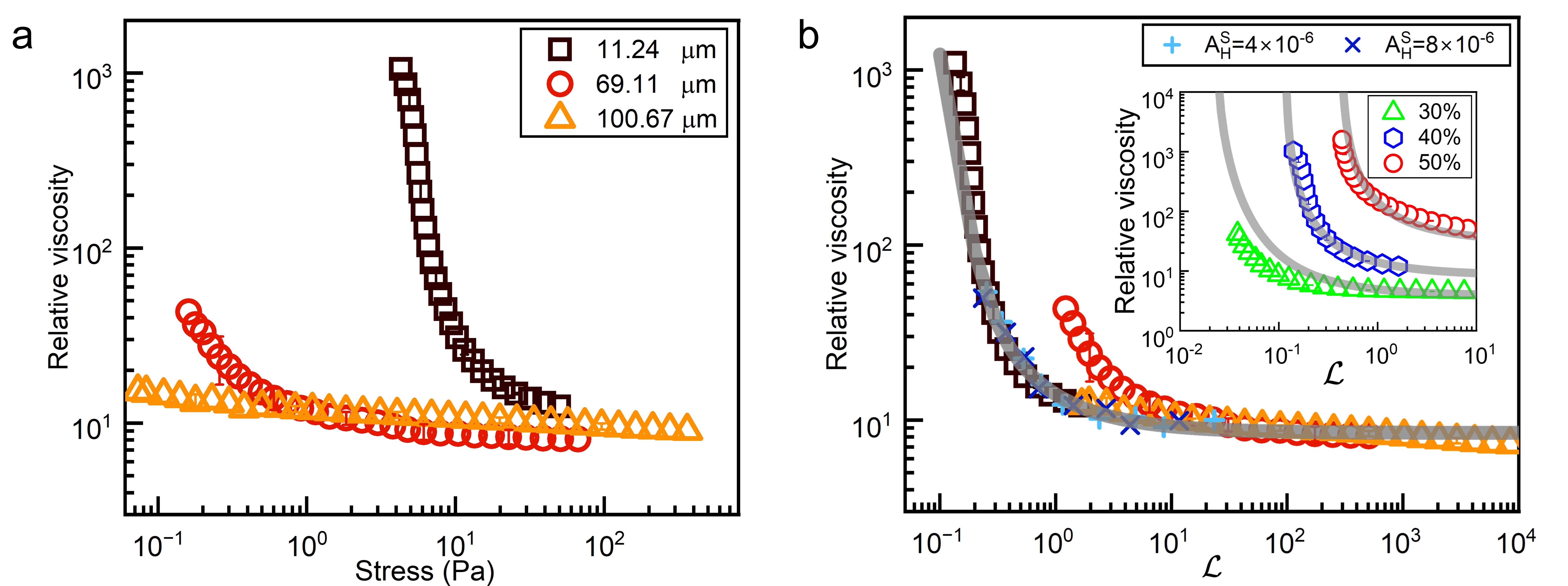}\\
  \caption{(a) \mdf{Evolution of the equilibrium viscosity with shear stress considering the changing of the particle size. (b) Viscosity curves re-scaled by $\mathcal{L}(a)$ number, in which the simulation result (cross and plus symbols) considering different $A_H^S$ and with $\epsilon= 10^{-3}a$, are included for comparison. The inset in (b) presents the fitting of the proposed model based on the $\mathcal{L}$ number for suspensions with changing particle volume fraction.} }\label{Fig3}
\end{figure}



Based on the Wyart and Cates (WC) model~\cite{Wyart2014,Richards2020} and the proposed $\mathcal {L}(a)$ number, we also {estimate} the fraction of particle in contact (particles in the organized structure) as,
\begin{equation}
\alpha(\mathcal{L}) = 1-e^{-\delta \mathcal{L}^{-\kappa}},
\end{equation}
where {$\delta =0.2238$ is} a correction factor that could be determined experimentally, {and $\kappa = 0.72$ determines the rate of breaking of the structure due to increasing shear rate.} Considering the condition of frictional {(constraining sliding)} and adhesive {(constraining rolling)} particles in the suspension~\cite{Richards2020}, the relative viscosity of the meso-scaled suspension with respect to the fluid medium {is} predicted as, $\eta_r = (1-\phi/\phi_j)^{-2}$, in which $\phi_j = \alpha\phi_{alp} + (1-\alpha)\phi_{\mu}$, with the adhesive loosing packing volume fraction, $\phi_{alp} \approx 0.29$, and the frictional jamming volume fraction, $\phi_{\mu} \approx 0.61$, respectively. As shown in Fig.~\ref{Fig3} (b), the master curve could be well described by the model based on the proposed $\mathcal{L}$ number, taking into account also the variation of the particle volume fraction.



\section{Conclusions}
To summarize, we {show} in this work that the meso-scaled particle suspension {is} strongly non-Newtonian due to the residual surface-type colloidal interaction such as the van der Waals interaction, and the behavior could be unified by the dimensionless $\mathcal{L}$ number based on a characteristic stress estimating the van der Waals interaction. More generally, non-Newtonian behaviors that are from particle interactions, e.g., the DLVO interactions, could be unified through the $\mathcal{L}$ number by accounting for the corresponding inter-particle interaction. As an example that {is} recently noted, for the system of clay dispersion, in which the strong electrostatic interactions {between clay particles} are taking into account, viscosity curves {is} unified with $\tau_p = \tau_y$, where $\tau_y$ {is} the yield stress that {measures} mainly the electrostatic inter-particle interaction~\cite{Lin2021c}. Therefore, the $\mathcal{L}$ proposed here could unify the non-Newtonian behavior of suspensions once the key interaction between the dispersion phase {is} identified. This offers an effective way to capture the rheological property of the multi-phase system. 

\begin{acknowledgments}
This work was supported by the National Natural Science Foundation of China (Grant No. 12222211).
\end{acknowledgments}

\bibliography{bibliography}

\end{document}